\documentstyle[sprocl]{article}

\input{psfig}
\newcommand{\nc}{\newcommand}

\nc{\npps}[3]{{\it  Nucl.\ Phys.\ Proc.\ Suppl.\ }{{\bf #1} {(#2)} {#3}}}
\nc{\advp}[3]{{\it  Adv.\ in\ Phys.\ }{{\bf #1} {(#2)} {#3}}}
\nc{\annp}[3]{{\it  Ann.\ Phys.\ (N.Y.)\ }{{\bf #1} {(#2)} {#3}}}
\nc{\apl}[3]{{\it  Appl. Phys. Lett. }{{\bf #1} {(#2)} {#3}}}
\nc{\apj}[3]{{\it  Ap.\ J.\ }{{\bf #1} {(#2)} {#3}}}
\nc{\apjl}[3]{{\it  Ap.\ J.\ Lett.\ }{{\bf #1} {(#2)} {#3}}}
\nc{\app}[3]{{\it Astropart.\ Phys.\ }{{\bf #1} {(#2)} {#3}}}
\nc{\cmp}[3]{{\it  Comm.\ Math.\ Phys.\ }{{ \bf #1} {(#2)} {#3}}}
\nc{\cqg}[3]{{\it  Class.\ Quant.\ Grav.\ }{{\bf #1} {(#2)} {#3}}}
\nc{\epl}[3]{{\it  Europhys.\ Lett.\ }{{\bf #1} {(#2)} {#3}}}
\nc{\ijmp}[3]{{\it Int.\ J.\ Mod.\ Phys.\ }{{\bf #1} {(#2)} {#3}}}
\nc{\ijtp}[3]{{\it Int.\ J.\ Theor.\ Phys.\ }{{\bf #1} {(#2)} {#3}}}
\nc{\jmp}[3]{{\it  J.\ Math.\ Phys.\ }{{ \bf #1} {(#2)} {#3}}}
\nc{\jpa}[3]{{\it  J.\ Phys.\ A\ }{{\bf #1} {(#2)} {#3}}}
\nc{\jpc}[3]{{\it  J.\ Phys.\ C\ }{{\bf #1} {(#2)} {#3}}}
\nc{\jap}[3]{{\it J.\ Appl.\ Phys.\ }{{\bf #1} {(#2)} {#3}}}
\nc{\jpsj}[3]{{\it J.\ Phys.\ Soc.\ Japan\ }{{\bf #1} {(#2)} {#3}}}
\nc{\lmp}[3]{{\it Lett.\ Math.\ Phys.\ }{{\bf #1} {(#2)} {#3}}}
\nc{\mpl}[3]{{\it  Mod.\ Phys.\ Lett.\ }{{\bf #1} {(#2)} {#3}}}
\nc{\ncim}[3]{{\it  Nuov.\ Cim.\ }{{\bf #1} {(#2)} {#3}}}
\nc{\np}[3]{{\it  Nucl.\ Phys.\ }{{\bf #1} {(#2)} {#3}}}
\nc{\pr}[3]{{\it Phys.\ Rev.\ }{{\bf #1} {(#2)} {#3}}}
\nc{\pra}[3]{{\it  Phys.\ Rev.\ A\ }{{\bf #1} {(#2)} {#3}}}
\nc{\prb}[3]{{\it  Phys.\ Rev.\ B\ }{{{\bf #1} {(#2)} {#3}}}}
\nc{\prc}[3]{{\it  Phys.\ Rev.\ C\ }{{\bf #1} {(#2)} {#3}}}
\nc{\prd}[3]{{\it  Phys.\ Rev.\ D\ }{{\bf #1} {(#2)} {#3}}}
\nc{\prl}[3]{{\it Phys.\ Rev.\ Lett.\ }{{\bf #1} {(#2)} {#3}}}
\nc{\pl}[3]{{\it  Phys.\ Lett.\ }{{\bf #1} {(#2)} {#3}}}
\nc{\prep}[3]{{\it Phys.\ Rep.\ }{{\bf #1} {(#2)} {#3}}}
\nc{\prsl}[3]{{\it Proc.\ R.\ Soc.\ London\ }{{\bf #1} {(#2)} {#3}}}
\nc{\ptp}[3]{{\it  Prog.\ Theor.\ Phys.\ }{{\bf #1} {(#2)} {#3}}}
\nc{\ptps}[3]{{\it  Prog\ Theor.\ Phys.\ suppl.\ }{{\bf #1} {(#2)} {#3}}}
\nc{\physa}[3]{{\it  Physica\ A\ }{{\bf #1} {(#2)} {#3}}}
\nc{\physb}[3]{{\it  Physica\ B\ }{{\bf #1} {(#2)} {#3}}}
\nc{\phys}[3]{{\it Physica\ }{{\bf #1} {(#2)} {#3}}}
\nc{\rmp}[3]{{\it  Rev.\ Mod.\ Phys.\ }{{\bf #1} {(#2)} {#3}}}
\nc{\rpp}[3]{{\it Rep.\ Prog.\ Phys.\ }{{\bf #1} {(#2)} {#3}}}
\nc{\sjnp}[3]{{\it Sov.\ J.\ Nucl.\ Phys.\ }{{\bf #1} {(#2)} {#3}}}
\nc{\spjetp}[3]{{\it Sov.\ Phys.\ JETP\ }{{\bf #1} {(#2)} {#3}}}
\nc{\yf}[3]{{\it Yad.\ Fiz.\ }{{\bf #1} {(#2)} {#3}}}
\nc{\zetp}[3]{{\it Zh.\ Eksp.\ Teor.\ Fiz.\  }{{\bf #1}  {(#2)} {#3}}}
\nc{\zp}[3]{{\it Z.\ Phys.\ }{{\bf #1} {(#2)} {#3}}}
\nc{\ibid}[3]{{\sl ibid.\ }{{\bf #1} {#2} {#3}}}

\def\be{\begin{equation}}
\def\ee{\end{equation}}
\def\bea{\begin{eqnarray}}
\def\eea{\end{eqnarray}}

\nc{\renc}{\renewcommand}

\def\lsim{\; \raise0.3ex\hbox{$<$\kern-0.75em
      \raise-1.1ex\hbox{$\sim$}}\; }
\def\gsim{\; \raise0.3ex\hbox{$>$\kern-0.75em
      \raise-1.1ex\hbox{$\sim$}}\; }

\nc{\cL}{{\cal L}}
\nc{\nn}{\nonumber \\*}
\def\GeV{{\rm\ GeV}}
\nc{\G}{\rm \ G}

\begin{document}
\hfill HIP-1998-40/th\vskip10pt
\title{Q-BALLS IN THE MSSM\footnote{Talk given at the PASCOS-98 meeting.}
}

\author{KARI ENQVIST}

\address{Department of Physics and Helsinki Institute of Physics, 
P.O. Box 9, FIN-00014 University
of Helsinki, Finland}


\maketitle
\abstracts{In the MSSM with gravity mediated supersymmetry breaking, there
may exist unstable but long-lived solitons carrying large baryonic charge, 
or B-balls. These decay well after
the electroweak phase transition, giving rise to B-ball baryogenesis.
Being made of squarks, B-ball decays produce also LSPs and hence
can be the source for all cold dark matter.}
%

A Q-ball is a stable, charge Q soliton in a scalar field theory 
with a spontaneously broken global $U(1)$ symmetry \cite{cole}.
The  Q-ball solution arises provided the scalar potential $V(\phi)$
is such that $V(\phi)/\vert\phi\vert^2$ has a minimum at non-zero $\phi$.
In the MSSM such solitons are necessarily part of the spectrum
of the theory, as was first pointed out by Kusenko \cite{k}. This is due
to the form of the MSSM scalar potential and the fact that squarks
and sleptons carry a global  $B-L$ charge. The crucial question
is whether B-balls and/or
L-balls can be copiously created  in the early Universe and whether
 they  could naturally be sufficiently long-lived to have 
important consequences for cosmology. 

An interesting possibility for Q-ball formation is provided 
by the fact that there are many flat directions in the MSSM \cite{drt}.
During inflation the MSSM scalar fields are free to fluctuate along
these flat directions and to form condensates. This is closely related 
to Affleck-Dine (AD) baryogenesis \cite{ad}, in which a 
B violating scalar condensate forms along a D-flat direction of the
 MSSM scalar potential composed of squark and possibly of slepton fields. 
The difference now is that 
the condensate is naturally unstable with respect
 to the formation of Q-balls \cite{cole,cole2,k} with a very large charge
\cite{ks}.

The properties  of the MSSM Q-balls will depend  upon 
the scalar potential
 associated with the condensate scalar, which in turn depends upon 
the SUSY breaking mechanism and on the order d
at which the non-renormalizable
terms lift the degeneracy of the potential; examples are the 
$H_{u}L$-direction with d=4 and $u^{c}d^{c}d^{c}$-direction
with d=6. If
SUSY breaking occurs at low energy scales, via gauge
 mediated SUSY breaking \cite{gmsb}, Q-balls will be stable \cite{ks,dks}. 
Stable B-balls 
could have a wide range of astrophysical, 
experimental and practical 
implications\cite{dks,ksdm}. 

A more conservative possibility is that SUSY 
breaking occurs via the supergravity hidden sector\cite{nilles}. 
In this case the potential
is not flat, but nevertheless radiative corrections to the $\phi^{2}$-type
 condensate potential allow B-balls to form \cite{qb1,qb2}.
Such B-balls can decay at temperatures less than
 that of the electroweak phase transition, $T_{ew}$, and consequently
 they could have important implications 
for baryogenesis. 

A particularly promising case is the d=6 $u^cu^cd^c$ flat direction \cite{qb1},
along which the potential reads
\be
V_6\simeq m^{2}_S |\phi|^{2} 
+ \frac{\lambda^{2}|\phi|^{10}
}{M_{p}^{6}} + \left( \frac{A_{\lambda} 
\lambda \phi^{6}}{M_{p}^{3}} + h.c.\right)    
~,\ee
where $\lambda$ and $A$ are coupling constants and 
the SUSY breaking mass $m_S^2\simeq m_0^2[1+K\log(\vert\phi\vert^2/\phi_0^2)]$,
where $\phi_0$ is the reference point and $K$ a negative constant (and
mainly due to gauginos), decreases as $\phi$ grows, thus allowing
B-balls to form. The potential is stabilized by the non-renormalizable term
so that the condensate field takes the value $\phi_0\simeq
4\times 10^{14}$ GeV. The decreasing of the effective mass term is
also responsible for the growth of any initial perturbation. In particular,
there are perturbations in the condensate field inherited from the
inflationary period. As was discussed in ref. 11, 
these will grow and become non-linear when
$H\simeq 2\vert K\vert m_S\alpha^{-1}$,
where $\alpha\simeq -log(\delta\phi_0(\lambda_0)/\phi_0)$
with $\lambda_{0}$  the length scale of the perturbation at $H \approx m_S$,
and $\phi_{0}$ is the value of $\phi$ when the condensate oscillations begin.
The charge of the condensate lump is determined by the baryon asymmetry 
of the Universe at $H_{i}$ 
and the initial size of the perturbation when it goes non-linear. 
The size of the initial non-linear region can also be determined \cite{qb1,qb2}.
The baryon asymmetry of the Universe at a given value of $H$ during 
inflaton oscillation
domination is given by 
\be n_{B} = \left( \frac{\eta_{B}}{2 \pi} \right) 
\frac{H^{2} M_{Pl}^{2}}{T_{R}}  \approx 
1.6 \times 10^{18} H^{2} \left(\frac{10^{9}}{T_{R}}\right)     
~,\ee
where we have taken the baryon to entropy ratio to be $\eta_{B} \simeq 10^{-10}$. 
It then follows that the charge in the initial condensate lump is given by 
\be B = \frac{4 \pi^{3}}{3 \sqrt{2}} 
\frac{\eta_{B} |K|^{1/2} M_{pl}^{2}}{m_S \alpha^{2} T_{R}} 
= 2 \times10^{15} |K|^{1/2} 
\left(\frac{100 \GeV}{m_S}\right)
\left(\frac{10^{9} \GeV}{T_{R}}\right)
\left(\frac{40}{\alpha}\right)^{2}
~\ee
where we have used $\alpha(\lambda_{0}) = 40$ as a typical value.

The formation of B-balls fromn the AD condensate can be shown to be 
generally effective \cite{qb2} if the charge density inside the initial lump is
small enough; this can be translated to a condition on the reheating
temperature which reads
\be  T_{R} \gsim \frac{\eta_{B} m M_{Pl}^{2}}{8 \pi \phi_{0}^{2}}
= 0.23 \left(\frac{m_S}{100 \GeV}\right)
 ~.\ee

After the formation B-balls could be dissociated by the bombardment of
thermal particles, or dissolve by charge escaping from the outer layers.
Both problems can be avoided \cite{qb2} 
provided $T_{R} \lsim 10^3-10^5$ GeV for 
$\vert K\vert$ in the range 0.01 to 0.1. It then follows from Eq. (5) that
the surviving B-balls must have very large charges, $B\gsim 10^{14}$. 
The decay rate of the
B-ball also depends on its charge;
it has been estimated to have an upper bound, 
which is likely to be saturated for B-balls 
with $\phi_{0}$ much larger than $m_{0}$ (as in the case of d=6 B-balls), and is 
given by \cite{cole2} 
\be
\frac{dB}{dt} = -f_s\frac{\omega^{3}A}{192 \pi^{2}}   
~,\ee
where $A$ is the area of the B-ball, 
$\omega \approx m_{0}$ for $|K|$ small compared with 1, and
$f_s$ is a possible enhancement factor due to condensate decaying into
scalars. We expect that $f_s\lsim 10^3$, although it is quite possible
that the B-ball is made of the lightest squark, in which case only
$f_s=1$. In the latter case the resulting
decay temperature is depicted in the Figure as a function of the Q-ball charge 
for both thin and thick-wall Q-balls, which have different surface areas
(d=6 B-ball is of the thick-wall variety\cite{qb1}).  As can be seen, for 
$B\gsim 10^{14}$, B-balls will indeed decay well below the electroweak
phase transition temperature, providing a new source of baryon
asymmetry not washed away by sphaleron interactions.
The only requirement is relatively low reheating temperature after inflation,
typically of the order of 1 GeV, which is fixed by the 
observed baryon asymmetry when the CP violating phase responsible for the 
baryon asymmetry is of the order of 1 \cite{qb2}. 
In particular, this is expected to be true for 
D-term inflation models \cite{kmr,bbbd}.

\begin{figure}
\leavevmode
\centering
\vspace*{32mm} 
\includegraphics{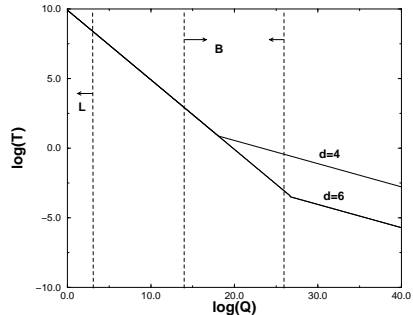}   
\caption{Q-ball decay temperature $T$ vs. the charge $Q$ for d=4 and d=6
Q-balls. The regions
where L-balls and B-balls exist are also indicated.}
\label{kuva2}       
\end{figure} 

Being made of 
squarks, B-balls initially decay to LSPs and baryons with a similar
number density (with three units of R-parity produced for each unit of
baryon number). Given the efficiency, $f_B$, by which B-balls are
created from the collapsing AD condensate, the B-ball produced LSP 
and baryonic densities will be related by
\be
\Omega_{BB} = {3 f_{B}m_{\rm LSP}\over m_N} \Omega_B    ~.
\ee
If the reheating temperature is smaller than the the freeze-out
temperature of the LSPs, given by $T_f\simeq m_{\rm LSP}/20$,
then B-balls would be the only source for cold dark matter. If 
$T_R\gsim T_f$, a relic LSP density would exist, with interesting
implications for the sparticle spectrum \cite{qbdm}.
Hence B-balls promise to provide a novel and complete alternative
to the more conventional cosmological scenario.

\section*{Acknowledgments}
I should like to thank John McDonald for many useful discussions on Q-balls
and for an enjoyable collaboration. This work has been supported by the
Academy of Finland.

\end{document}